\documentstyle[epsf, rotate]{elsart}
\begin{document}

\begin{frontmatter}
\title{An Unusual Antagonistic Pleiotropy in the Penna Model for 
Biological Ageing}

\author{\bf A.O. Sousa\thanksref{email1}} 
\author{\bf and S. Moss de Oliveira\thanksref{email2}}
 
\address{\it Instituto de F\'{\i}sica, Universidade Federal 
Fluminense \\ Av. Litor\^anea s/n, Boa Viagem, 24210-340 Niter\'oi, RJ, 
Brazil}
\thanks[email1]{e-mail:sousa@if.uff.br}
\thanks[email2]{e-mail:suzana@if.uff.br}

\date{\today}
\maketitle

\begin{abstract}
We combine the Penna  Model for biological aging,
which is based on the mutation-accumulation theory, with a sort of
antagonistic pleiotropy. We show that depending on how the 
pleiotropy is introduced, it is possible to reproduce both the 
humans mortality, which increases exponentially with 
age, and fruitfly mortality, which decelerates at old ages, 
allowing the appearance of arbitrarily old Methuselah's.
\end{abstract}

\begin{keyword}
Ageing. Penna model. Monte Carlo Simulations.
\end{keyword}

\end{frontmatter}
%\pacs{PACS numbers: 05.50.+q, 02.70.Lq, 75.10.Hk}
\section{Introduction}

According to the evolutionary theory of aging the ultimate cause of 
senescence is the declining force of natural selection with age
\cite{rose}. 
From 1950s to 1980s, this idea has been formulated in a
progressively more explicit and formal way, culminating in the
mathematical treatment  of Charlesworth \cite{charlesworth80}.
Several different genetic mechanisms may be involved in this lack
of selection pressure at old ages, but the two major alternative 
population genetic mechanisms are the antagonistic pleiotropy and 
the mutation-accumulation mechanism.

Pleiotropic genes are those that have different effects on fitness at 
different ages. The antagonistic pleiotropy mechanism 
\cite{medawar46,medawar52,williams57,williams60} proposes that
ageing is caused by pleiotropic genes that have opposite effects at 
young and old ages, that is, genes that confer some 
advantage during the period of maximum reproductive probability, at the
expenses of some disadvantage or declining vigor late in life. An
example is the accumulation of calcium to build up bones first, and 
the appearence of arteriosclerosis later. The
mutation-accumulation mechanism \cite{charlesworth80,medawar52,edneygill68} 
proposes that ageing is caused by the 
accumulation of deleterious alleles having effects that appear only at old ages, when 
selection pressure is already weak, and in this way may remain in the 
population, being transmitted from parents to offspring.

After the analytically treated mathematical model of Partridge and
Barton in 1993 \cite{pb}, a continuous flow of Monte Carlo simulations of
ageing started (for a review see \cite{stauffer94}). The bit-string
Penna model of life history was published in 1995 \cite{penna} and is 
now by far the most widely used Monte Carlo simulation technique to 
predict many of the features related to ageing (for a review see 
\cite{ourbook}). It is based on the mutation-accumulation theory and 
gives that the probability to die within the next
year increases exponentially with age, in agreement with the 19th
century empirical Gompertz law for human mortality.

In this paper we combine the Penna model with an antagonistic
pleiotropy mechanism. However, this antagonistic pleiotropy does not work 
in the usual way explained above, since it does not give any
advantage at young ages. On the contrary, it gives a disadvantage
before reproduction age and some advantage at old ages. Our biological 
motivation is the existence of a rather curious gene involved in the 
cholesterol transportation by the blood stream \cite{recherche}. This
gene can be
found in three different forms, $\epsilon 2$, $\epsilon 3$ and
$\epsilon 4$. The allele $\epsilon 4$ seems to be the most dangerous one: 
those who carry it have a high cholesterol level and a high
probability of suffering from heart diseases and also Alzheimer. In 
contrast, the allele $\epsilon 2$ seems to be relatively beneficial:
it is associated to low levels of cholesterol and a low incidence of 
Alzheimer. The allele $\epsilon 3$ has an intermediate influence
concerning these diseases. Surprisingly, studies in different
populations have shown that the form $\epsilon 3$ is 
by far the most common one, followed by the dangerous form
$\epsilon 4$. The less common form is the beneficial one, $\epsilon2$. 
Since deleterious mutations that may affect the individuals at 
young ages are generally eliminated from the population, the 
persistence of the allelic form $\epsilon4$ may be related to some 
antagonistic effect.

Combinations of the usual antagonistic pleiotropy mechanism with 
the Penna model can be found in \cite{bernardes,almeida}; simple pleiotropy, 
understood as a two-fold effect of a harmfully mutated gene, 
has also been recently introduced into the model in order to show 
the advantages of sexual reproduction \cite{staumartins}.

This paper is organized as follows: in section 2 we review the 
traditional sexual 
version of the Penna Model; the readers already familiar with the 
model can skip this section. In section 3 we present our pleiotropy 
mechanism and results, and in section 4, our conclusions.

\section{The Penna Model for Sexual Populations}

In the sexual version of the model, each individual of a population 
is represented by a
``chronological genome'' that consists of two bit-strings (diploid) of
32 bits (zeroes and ones). These two strings are read in
parallel, which means that each position is associated to two 
bits (alleles). One time step corresponds to reading a new bit position
in all genomes. The first position contains the information about the
individual's first period of life, the second one about the second
period and so on. Each individual can live at most for 32
periods, each period corresponding to one day, week or year depending
on the species.
Genetic defects (harmful mutations) are represented by bits 
1 and healthy genes by bits 0. Homozygous loci (positions) 
are those with two equal alleles
(bits). Heterozygous loci are those corresponding to
different alleles, meaning that the individual inherited opposite
informations, 0 and 1, from each parent. If an individual has two bits 1
(homozygous) at the third position, for instance, it starts to suffer
the effects of a genetic defect at its third period of life. If it is an
homozygous position with two bits zero, no disease appears at that
age. If the individual is heterozygous in some position, it will
become sick only if the mutation at this position has a dominant effect.
Six of the 32  bits positions are randomly chosen, at the
beginning of the simulation, where the alleles 1 are dominant; in the
remaining 26 positions this allele is recessive. These
dominant positions are the same for all individuals and kept fix 
during the whole process. When the number of accumulated diseases of
any individual reaches a threshold value $T$, the individual dies.

If a female succeeds in surviving until the minimum reproduction age
$R$, it generates, with probability 0.5, one offspring every period  
until death. The female
randomly chooses a male to mate, with age also greater or equal to
$R$. The offspring genome is constructed from the parents' ones;
first the strings of the mother are randomly crossed, and a female
gamete (one string of 32 bits) is produced. $M_f$ deleterious
mutations are then randomly
introduced. The same process occurs with the father's genome (with
$M_m$ mutations), and the union of the two remaining gametes form the
new genome. The sex of the baby is randomly chosen, each one with
probability $50\%$. 

Deleterious mutation means that if the randomly chosen bit of the
parent's gamete is equal to 1, it remains 1 in the offspring genome,
but if it is equal to zero in the parent's gamete, it is set to 1 in
the baby genome. The most important characteristic of this dynamics
is that the bits 1 accumulate, after many generations, at the end part
of the genomes, that is, after the minimum reproduction age $R$. For
this reason ageing appears: the survival probabilities per period 
decrease with age, in agreement with the mutation accumulation 
hypothesis and reality \cite{salmon}.

Although only harmful mutations are considered, a population that
increases exponentially in time is obtained. In order to avoid this
exponential increase, each individual has also a 
probability to die due to a lack of food or space. This probability  
is given by the Verhulst factor
$$V=N(t)/N_{max} \,\,\, ,$$ where  $N(t)$ is the number of individuals
at time $t$ and $N_{max}$ is the maximum carrying capacity of the
environment, which is defined at the beginning of the simulation. 
At every time step, for each individual, a
random number between 0 and 1 is generated and compared to $V$: if
this number is smaller than $V$ the individual dies, independently of
its age and genome. 

\section{Pleiotropy and Results}

Two different strategies of the antagonistic pleiotropy mechanism are
now introduced into the Penna model. In both cases we chose a priori two 
different positions
among the 32 possible ones, to be considered as special positions. Let
us call them $markage1$ and $markage2$. 
{\bf If, and only if,} an individual has two bits 1 (homozygous) at  
$markage1$, or if only one bit is equal to 1 but $markage1$ happens to 
be a dominant position, and the individual succeeds in surviving until 
the age $markage2$, then:  

\begin{itemize}
\item \noindent Case 1: There is a probability $P_{clean}$ that the 
current number of accumulated mutations is decreased by one.  
When the individual reaches 
$markage2$, a random number between 0 and 1 is generated. 
If it is greater or equal to $P_{clean}$, the number of accumulated
diseases is decreased. If it fails, in the next time step another
random number is generated, and the process is repeated until it works. 
So in this case the number of accumulated 
mutations is decreased by one {\bf only once}, at an age that can be 
equal or greater than $markage2$.
\bigskip

\item \noindent Case 2: Every year, from $age=markage2$ 
until death, there is a probability $P_{clean}$ that the 
current number of accumulated mutations is decreased by one. 
So in this second case there is a {\bf cumulative probability} 
of decreasing the current number of accumulated mutations.
\end{itemize}

It is important to notice that in both cases the genomes of the
individuals are not changed: only the counter of the number of 
mutations is decreased by one.  

We measure the survival rates and the mortalities of the populations
as a function of age. For an already stable population the survival 
rate is defined as the ratio between the number of individuals with 
age $a+1$ and the number of individuals with age $a$ : 
\begin{equation}
S(a) = \frac {N_{a+1}(t)} {N_a(t)} \, \, .
\end{equation}
The mortality is defined as \cite{racco}:
\begin{equation}
q(a) = - \ln \left [1 - \frac {D_a (t)} {N_a (t)} \right ] \, \, ,
\end{equation}
where $D_a$ is the number of deaths, due to genetic diseases, at age 
$a$.

\noindent The figures below were obtained using the following parameters:

\begin{itemize}
\item \noindent Initial population = $10^5$ individuals (half males
and half females) and maximum population size = $10^6$ individuals;
\item \noindent Minimum reproduction age $R=10$;
\item \noindent Limit number of accumulated diseases $T = 3$;
\item \noindent Mutation rates $M_m = M_f = 1$;
\item \noindent Pleiotropic positions $markage1 = \, 9$ and $markage2 
= \, 16$; 
\item \noindent Results averaged from 100 independent runs of $10^5$  
steps each. For each run measurements are averaged from the last 5000 steps.
\end{itemize}

In figure 1 we show the mortalities versus age in a 
linear-log scale, for the
traditional Penna model (circles) and for our case 1. Squares 
correspond to $P_{clean}=0.4$ and diamonds to $P_{clean}=0.5$. 
As can be seen this strategy gives the same results as the traditional 
model, with the curves presenting an ``s'' shape, as already
obtained before \cite{ourbook}. (In fact, a much better agreement with the
Gompertz law is obtained for asexual populations \cite{racco}, where the 
traditional Penna model gives a straight line for the linear-log plot 
of the mortality as a function of age, for ages above the minimum
reproduction age $R$).

In figure 2 we show the results for our case 2, where a strong
difference between the standard result (circles)   
and the cases where the cumulative probability $P_{clean}$ is adopted 
can be noticed.
Now the mortalities decelerate at older ages, as observed for
drosophilas \cite{curtsinger,promislow}, and 
may even present a peak (triangles), as observed for 
medflies \cite{carey}. (For a review on experimental results 
see \cite{zeus}). The small sharp peaks at age 9 (squares and 
diamonds), which 
corresponds to our $markage1$, shows that the population prefers to have 
the 9-th bit set to 1, in order to profit from a decrease of the
probability to die at older ages. 
   
In figure 3, we show the survival rates for the same cases presented 
in figure 2. Again it can be seen a decrease of the survival
probabilities at age 9, for $P_{clean}$ equal to 0.4 and 0.5, showing 
that after many generations there is a fixation of the 9-th bit inside 
the population.

\section{Conclusions}

We introduce an anusual form of antagonistic pleiotropy into the 
bit-string Penna model for biological ageing. We show that   
it is possible to simulate the drosophila mortality, which 
decelerates at old ages, as well as the medfly mortality, 
which presents a peak around a given age (Fig. 2), even considering only 
homogeneous stable populations. Depending on how the pleiotropy is 
introduced, the exponential increase of human mortality with age is 
also obtained (Fig. 1).

\section*{Acknowledgments}
The authors thank D. Stauffer, P.M.C. de Oliveira, T.J.P. Penna and 
J.S. S\'a Martins for useful discussions. This work is partially 
supported by the Brazilian agencies CAPES, CNPq and FAPERJ.

\newpage
\section*{Figure Captions}

Figure 1 - Mortalities as a function of age in a linear-log scale. 
Circles correspond to the traditional Penna model; squares and
diamonds correspond to our case 1, for $P_{clean} = 0.4$ and 
$P_{clean} = 0.5$, respectively. Other parameters are mentioned in 
section 3.
\bigskip

Figure 2 - Mortalities as a function of age, now for the cumulative probabilities
of case 2. Circles: traditional Penna model; 
Triangles: $P_{clean} = 0.2$; Squares: $P_{clean} = 0.4$; Diamonds: 
$P_{clean} = 0.5$.

\bigskip

Figure 3 - The survival rates as a function of age for the traditional 
model and for our case 2. The symbols are the same of Fig. 2. 
 
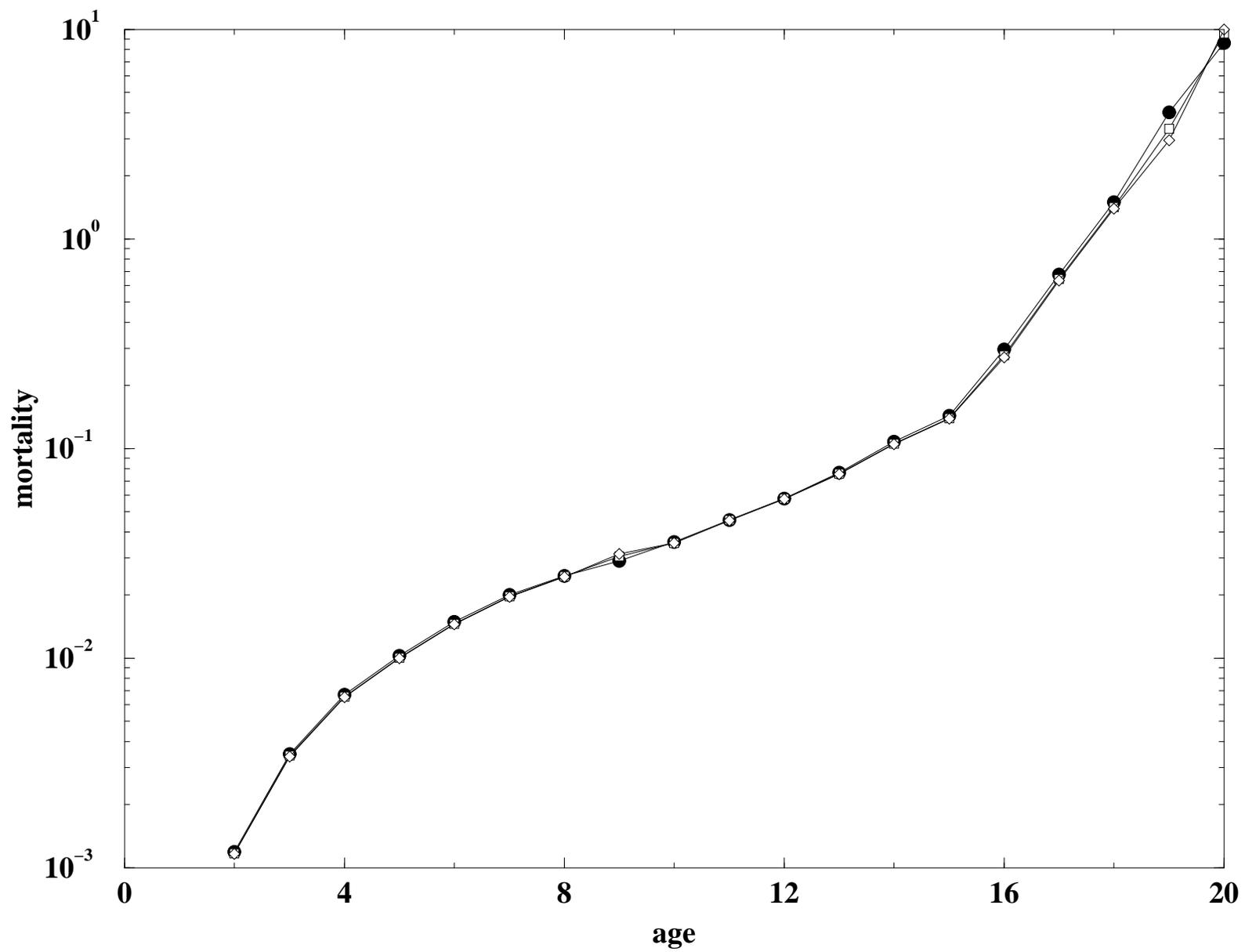
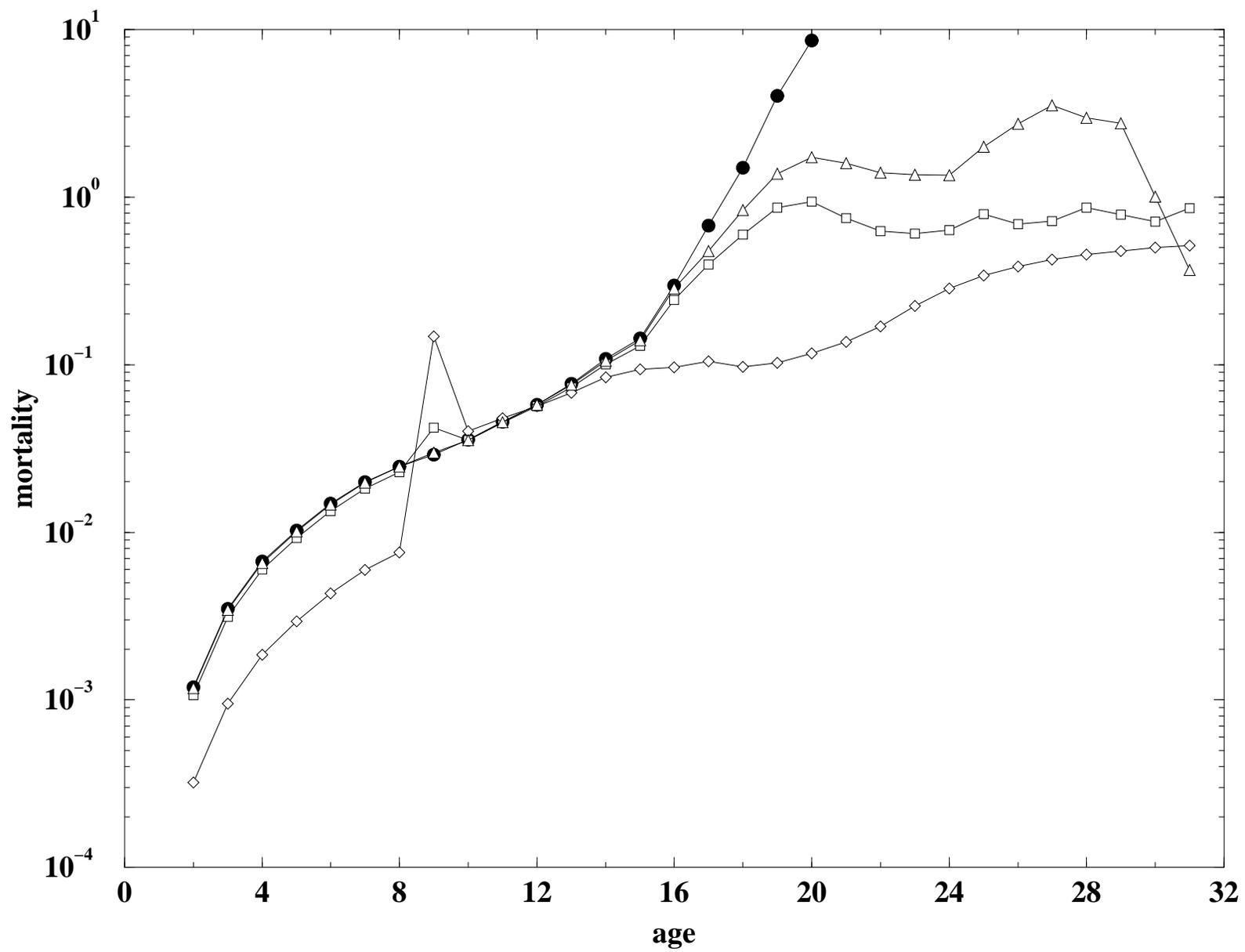
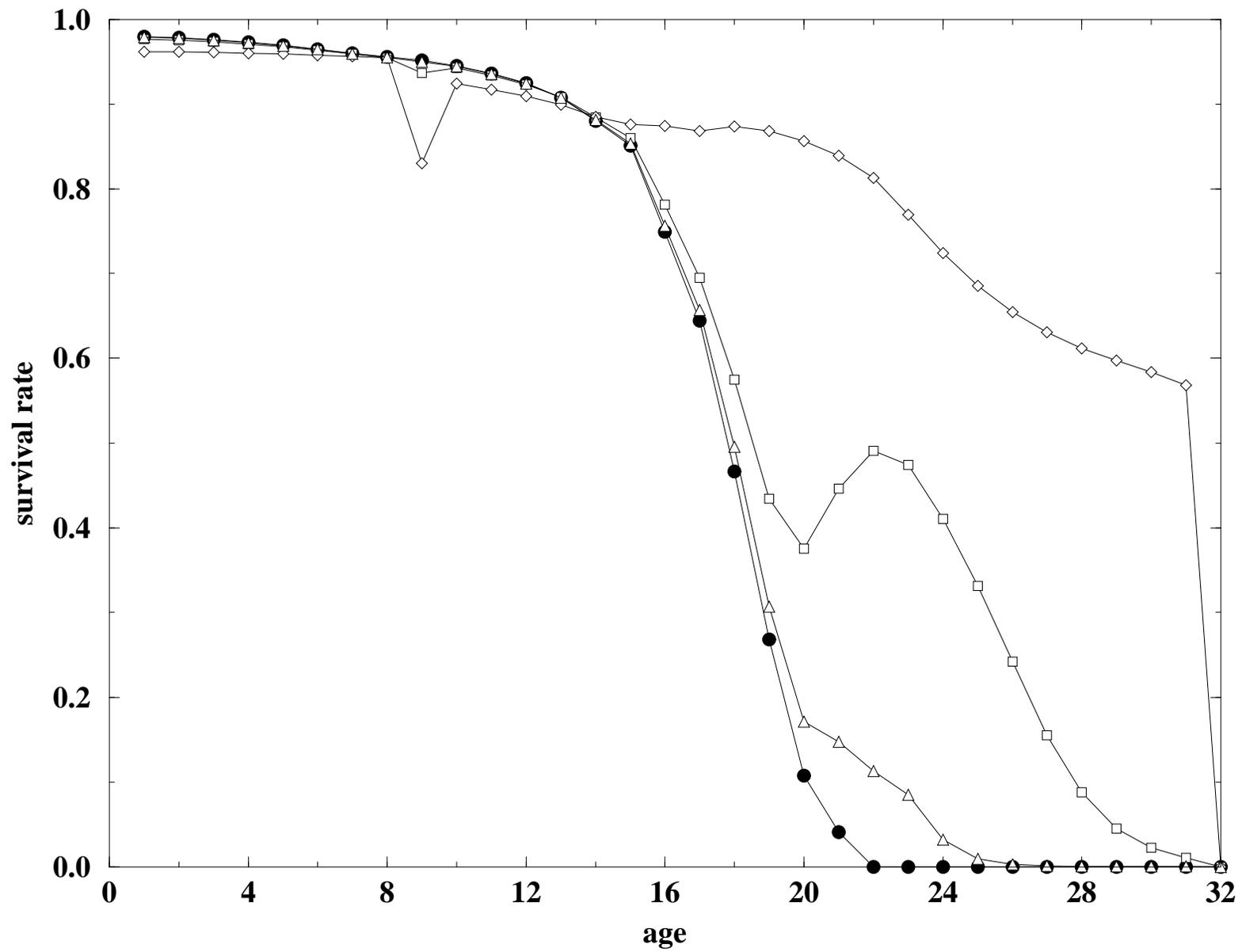
\end{document}